\documentclass[prd,showpacs,preprintnumbers,amsmath,amssymb,nofootinbib]{revtex4}

\usepackage{graphicx,color}
\usepackage{dcolumn}
\usepackage{bm}
\usepackage{dsfont,amsmath,amssymb}
\newcommand{\Od}{{\cal O}}

\newcommand{\tr}{\mbox{tr}}

\newcommand{\diag}{\mbox{diag}}

\newcommand{\condtwo}{\langle \bar q q \rangle}
\newcommand{\condtwoi}{\langle \bar q_i q_i \rangle}
\newcommand{\condu}{\langle \bar u u \rangle}
\newcommand{\condd}{\langle \bar d d \rangle}
\newcommand{\conds}{\langle \bar s s \rangle}
\newcommand{\condsum}{\langle \bar u u + \bar d d\rangle}
\newcommand{\conddif}{\langle \bar u u - \bar d d\rangle}

\newcommand{\gsim}{\raise.3ex\hbox{$>$\kern-.75em\lower1ex\hbox{$\sim$}}}
\newcommand{\lsim}{\raise.3ex\hbox{$<$\kern-.75em\lower1ex\hbox{$\sim$}}}

\begin{document}

\title{Isospin-Breaking quark condensates in Chiral Perturbation Theory}

\author{A. G\'omez Nicola}
\email{gomez@fis.ucm.es} \affiliation{Departamento de F\'{\i}sica
Te\'orica II. Univ. Complutense. 28040 Madrid. Spain.}
\author{R. Torres Andr\'es}
\email{rtandres@fis.ucm.es} \affiliation{Departamento de
F\'{\i}sica Te\'orica II. Univ. Complutense. 28040 Madrid. Spain.}

\begin{abstract}

We analyze the isospin-breaking corrections to quark condensates
  within one-loop $SU(2)$ and $SU(3)$  Chiral Perturbation Theory including  $m_u\neq m_d$  as well as
 electromagnetic (EM) contributions. The explicit expressions are given and several phenomenological aspects are studied. We analyze the sensitivity of recent condensate determinations to the EM
low-energy constants (LEC). If the explicit chiral symmetry breaking induced by EM terms generates a ferromagnetic-like response of the vacuum, as in the case of quark masses, the increasing of the
 order parameter implies constraints  for the EM LEC, which we check with different
estimates in the literature.  In addition, we extend  the sum rule
    relating quark condensate ratios in $SU(3)$ to include EM corrections, which are  of the same order as the $m_u\neq m_d$ ones, and we use that sum rule to estimate the vacuum asymmetry within ChPT. We also discuss the matching
conditions  between the $SU(2)$ and $SU(3)$ LEC involved in the
condensates, when both isospin-breaking sources are taken into account.
 \end{abstract}



\maketitle

\section{Introduction}
The low-energy sector of QCD has been successfully described over recent years within the chiral lagrangian framework. Chiral Perturbation Theory (ChPT)
is based on the spontaneous breaking of chiral symmetry $SU_L(N_f)\times SU_R(N_f)\rightarrow SU_V(N_f)$ with $N_f=2,3$ light flavours and provides a consistent and systematic model-independent scheme to calculate low-energy observables \cite{we79,Gasser:1983yg,Gasser:1984gg}. The effective ChPT lagrangian is constructed as the more general expansion  ${\cal L}={\cal L}_{p^2}+{\cal L}_{p^4}+\dots$ compatible with the QCD underlying symmetries, where $p$ denotes derivatives or meson mass and external momentum   below the chiral scale $\Lambda_{\chi}\sim$ 1 GeV.

 The $SU_V(N_f)$ group of vector transformations corresponds to the isospin symmetry for $N_f=2$. In the $N_f=3$ case, the vector group symmetry is broken by the strange-light quark mass difference $m_s-m_{u,d}$, although $m_s$ can still be considered as a perturbation compared to  $\Lambda_{\chi}$, leading to $SU(3)$ ChPT \cite{Gasser:1984gg}. In the $N_f=2$ case, the isospin symmetric limit is a very good approximation in Nature. However, there
 are several known examples where isospin breaking is phenomenologically relevant at low energies, such as sum
 rules
 for quark condensates \cite{Gasser:1984gg}, meson masses and corrections to Dashen's theorem \cite{Urech:1994hd},
 pion-pion \cite{Meissner:1997fa,Knecht:1997jw} and pion-kaon \cite{Nehme:2001wf,Kubis:2001ij} scattering  in connection with mesonic atoms
 \cite{Knecht:2002gz,Schweizer:2004qe},
 CP violation \cite{Ecker:1999kr}, $a_0-f_0$ mixing
 \cite{Hanhart:2007bd}, kaon decays \cite{Nehme:2004xf,Colangelo:2008sm}
 and other hadronic observables (see \cite{Rusetsky:2009ic} for a recent review).

 The two possible sources of isospin breaking  are  the $m_d-m_u$ light quark mass difference and electromagnetic (EM)  interactions.
 Both can be accommodated within the ChPT framework. The former is accounted for by modifying the quark mass matrix and generates a
 $\pi^0\eta$ mixing term in the $SU(3)$ lagrangian \cite{Gasser:1984gg}. The expected corrections from this source are of  order
 $(m_d-m_u)/m_s$. On the other hand,  EM interactions, which in particular  induce mass differences between charged and neutral
 light mesons,  can be included in  ChPT via the external source method and
 give rise to new terms in the effective lagrangian
 \cite{Ecker:1988te,Urech:1994hd,Knecht:1997jw,Meissner:1997fa,Neufeld:1995mu,Schweizer:2002ft}
 of order ${\cal L}_{e^2}$, ${\cal L}_{e^2p^2}$ and so on,with $e$ the electric charge. These terms fit into
  the ChPT power counting scheme by considering formally $e^2=\Od (p^2/F^2)$, with $F$ the pion decay constant in the chiral limit.

The purpose of this paper is to study the isospin-breaking
corrections to quark condensates, whose main importance is their
relation to the symmetry properties of the QCD vacuum. The singlet
contributions $\langle \bar u u + \bar d d\rangle$ for $SU(2)$ and
$\langle \bar u u + \bar d d+ \bar s s \rangle$ for $SU(3)$ are
 order parameters for chiral symmetry, while the isovector one
$\langle \bar u u - \bar d d\rangle$ behaves as an order parameter
for isospin breaking, which is not spontaneously broken
\cite{Vafa:1983tf}. We will calculate the condensates within
one-loop ChPT, which ensures the model independency of our
results, and will address several phenomenological consequences. The
two sources of isospin-breaking will be treated consistently on
the same footing, which will  allow us to test the sensibility of previous phenomenological analysis to the EM low-energy constants (LEC). Moreover, the EM corrections induce an explicit breaking
of  chiral symmetry which will lead to lower bounds
  for certain combinations of the LEC involved, provided the vacuum response is ferromagnetic, as in the case of quark masses. In addition, in  $SU(3)$  one can derive a
sum rule relating the different condensate ratios for $m_u\neq
m_d$ \cite{Gasser:1984gg} which, as we will show here,
receives an EM correction  not  considered before and of the same order as that proportional to $m_u-m_d$. The latter is useful to estimate the vacuum asymmetry $\condd/\condu$ reliably within ChPT.    An additional aspect that we will discuss is the matching of the $SU(2)$ and $SU(3)$ LEC combinations appearing in the condensates when both isospin-breaking sources are present, comparing with previous results in the literature. The analysis carried out in the present work will serve also to establish a firm phenomenological basis for its extension  to finite temperature, in order to study different aspects related to chiral symmetry restoration \cite{Nicola:2011gq}.

    With the above motivations in mind, the paper is organized as follows: in section \ref{sec:form} we
briefly review  the effective lagrangian formalism needed for our
present work, paying special attention to several theoretical
issues and to the  numerical values of the parameters and LEC
needed here. Quark condensates for $SU(2)$ are calculated and
analyzed in section \ref{sec:condsu2}, where we discuss the
general aspects of the bounds for the EM LEC based on chiral
symmetry breaking. In that section we also comment on the analogy
with lattice analysis.  The $SU(3)$ case is separately studied in
section \ref{sec:condsu3}. In that section, we perform first
a numerical analysis of the isospin-breaking corrections, paying
special attention to  the effect of the EM LEC in  connection with
previous  results in the literature. In addition, we obtain the EM corrections to the sum
rule for condensate ratios, which we use to estimate the vacuum
asymmetry within ChPT. We also provide
the  LEC bounds for this case, checking them with previous LEC
estimates and, finally, we discuss the matching conditions for the LEC involved.   In Appendix \ref{app:lag} we collect the
lagrangians of fourth order and the renormalization
 of the LEC  used in the main text.

\section{Formalism: effective lagrangians for isospin breaking}
\label{sec:form}

 The effective chiral lagrangian up to fourth order  is given schematically by

\begin{equation}
{\cal L}_{eff}={\cal L}_{p^2+e^2}+{\cal L}_{p^4+e^2p^2+e^4}.
\label{efflag}
\end{equation}
The second order lagrangian is the familiar non-linear sigma model, including now the gauge coupling of mesons to the EM field through the covariant derivative, plus an extra  term:

\begin{equation}\label{L2}
\begin{split}
{\cal L}_{p^2+e^2} &=\frac{F^2}{4} \tr\left[D_\mu U^\dagger D^\mu U+2B_0{\cal M}\left(U+U^\dagger\right)\right] \\
&\quad +C\tr\left[QUQU^\dagger\right].
\end{split}
\end{equation}

Here, $F$ is the pion decay constant in the chiral limit and $U(x)\in SU(N_f)$ is the Goldstone Boson (GB)
field in the exponential representation
$U=\exp [i\Phi/F]$ with :
\begin{eqnarray}
SU(2): \Phi&=& \left(\begin{array}{ll}
 \pi^0 &  \sqrt{2}\pi^+ \\
 \sqrt{2}\pi^- & - \pi^0
    \end{array}\right),\nonumber\\
SU(3): \Phi&=&\left(\begin{array}{lll}
 \pi^0+\frac{1}{\sqrt{3}}\eta& \sqrt{2}\pi^+ & \sqrt{2}K^+ \\
\sqrt{2}\pi^- & -\pi^0+\frac{1}{\sqrt{3}}\eta & \sqrt{2} K^0 \\
 \sqrt{2}K^-& \sqrt{2}\bar K^0 & \frac{-2}{\sqrt{3}}\eta
    \end{array}\right),
\end{eqnarray}
with $\eta$ the octet member with $I_3=S=0$. The covariant
derivative is $D_\mu=\partial_\mu+iA_\mu[Q,\cdot]$ with $A$ the EM
field. ${\cal M}$ and $Q$ are the quark mass and charge matrices,
i.e., in $SU(3)$ ${\cal M}=\diag (m_u,m_d,m_s)$ and
$Q=\diag(e_u,e_d,e_s)$ with $e_u=2e/3,e_d=e_s=-e/3$ for physical
quarks. The additional term in (\ref{L2}), the one proportional to
$C$, can be understood as follows: the QCD lagrangian for
$m_u=m_d$ coupled to the EM field is not invariant
under an isospin transformation $q\rightarrow g q$ with $g\in
SU(N_f)$ and $q$ the quark field. However, it would be isospin
invariant if the quark matrix $Q$ is treated as an external field
transforming as $Q\rightarrow g^\dagger Q g$.  Therefore, the
low-energy effective lagrangian has to include all possible terms
compatible with this new symmetry, in addition to the standard QCD
symmetries. The lowest order $\Od(e^2)$ is the $C$-term in
(\ref{L2}), since $U$ transforms as $U\rightarrow g^\dagger U g$.
Actually, one allows for independent ``spurion" fields $Q_L(x)$
and $Q_R(x)$ transforming under  $SU_L(N_f)\times SU_R(N_f)$ so
that  one can  build up the new  possible terms to any order in
the chiral lagrangian expansion, taking in the end $Q_L=Q_R=Q$
\cite{Urech:1994hd}.

In the previous expressions,  $F,B_0 m_{u,d,s}, C$ are the low-energy parameters to this order. Working out the kinetic terms, they can be directly related to the leading-order tree level values for the  decay constants  and masses of the GB. In  $SU(2)$, the tree level masses to leading order are:
\begin{eqnarray}
M_{\pi^+}^2&=&M_{\pi^-}^2=2\hat m B_0 + 2 C \frac{e^2}{F^2},\nonumber\\
M_{\pi^0}^2&=& 2\hat m B_0, \label{treemassessu2}
\end{eqnarray}
with $\hat m=(m_u+m_d)/2$ the average light quark mass. Note that both terms contributing to the charged pion mass are of the same order in the chiral power counting, although numerically $\left(M_{\pi^\pm}^2-M_{\pi^0}^2\right)/M_{\pi^0}^2\simeq 0.1$, which we will use in practice  as a further perturbative parameter to simplify some of the results.

In the $SU(3)$ case,  the mass term in (\ref{L2}) induces a mixing contribution between the $\pi^0$ and the $\eta$ meson fields given by
${\cal L}_{mix}=(B_0/\sqrt{3})(m_d-m_u)\pi^0\eta$. Therefore, the
kinetic term has to be brought to the canonical form before
identifying the GB masses, which is performed by the field
rotation \cite{Gasser:1984gg}:

\begin{eqnarray}
\pi^0&=&\bar\pi^0\cos\varepsilon-\bar\eta\sin\varepsilon,\nonumber\\
\eta&=&\bar\pi^0\sin\varepsilon+\bar\eta\cos\varepsilon,
\label{pietarot}
\end{eqnarray}
where the mixing angle is given by:

\begin{equation}
\tan 2\varepsilon=\frac{\sqrt{3}}{2}\frac{m_d-m_u}{m_s-\hat m}.
\label{mixangle}
\end{equation}

Once the above $\pi^0\eta$ rotation is carried out, the $SU(3)$
tree level meson masses to leading order read:
\begin{eqnarray}
M_{\pi^+}^2&=&M_{\pi^-}^2=2\hat m  B_0 + 2 C \frac{e^2}{F^2},\nonumber\\
M_{\pi^0}^2&=&  2B_0\left[\hat m -\frac{2}{3} (m_s-\hat m)\frac{\sin^2\varepsilon}{\cos 2\varepsilon}\right],\nonumber\\
M_{K^+}^2&=&M_{K-}^2=(m_s+m_u) B_0 + 2 C \frac{e^2}{F^2},\nonumber\\
M_{K^0}^2&=&(m_s+m_d) B_0,\nonumber\\
M_\eta^2&=&2B_0\left[\frac{1}{3}(\hat m + 2m_s)+\frac{2}{3}(m_s-\hat m)\frac{\sin^2\varepsilon}{\cos 2\varepsilon}\right].
\label{treemassessu3}
\end{eqnarray}

The above five equations are the extension of the Gell-Mann-Oakes-Renner (GOR) relations \cite{GellMann:1968rz} to the isospin asymmetric case and  allow  to relate the four constants
$B_0 m_{u,d,s}$ and $C$ ($\varepsilon$ is given in terms of  quark masses in (\ref{mixangle})) with the five meson masses or combinations of them.
The additional equation
 provides the following relation between the tree level LO masses:

\begin{equation}
\left(M_{K^\pm}^2-M_{\pi^\pm}^2\right)^2-3\left(M_\eta^2-M_{K^0}^2\right)\left(M_{K^0}^2-M_{\pi^0}^2\right)=0,
\label{gmoext}
\end{equation}
The above equation is compatible with the one obtained in \cite{Ditsche:2008cq} neglecting $\Od(m_u-m_d)^2$ terms. Actually, note that although all terms in (\ref{treemassessu3}) are formally of the same chiral order, numerically (see below) we expect $\varepsilon\sim (\sqrt{3}/4)(m_d-m_u)/m_s\ll 1$ and hence the mixing-angle corrections to the squared masses  to be $\Od(M_\pi^2\varepsilon)$ and $\Od(M_\eta^2\varepsilon^2)$ for the neutral pion and eta respectively. On the other hand, in the isospin-symmetric limit ($m_u=m_d$ and $e=0$) (\ref{gmoext}) is nothing but the Gell-Mann-Okubo
  formula $4M_K^2-3M_\eta^2-M_\pi^2=0$ \cite{GMO}. Neglecting only the $m_d-m_u$ mass difference in (\ref{treemassessu3}) leads to Dashen's theorem $M_{K^\pm}^2-M_{K^0}^2=M_{\pi^\pm}^2-M_{\pi^0}^2$ \cite{Dashen:1969eg} and then
  eq.(\ref{gmoext}) reduces to $4M_{K^0}^2-3M_\eta^2-M_{\pi^0}^2=0$, i.e., the Gell-Mann-Okubo formula for neutral states.
 However, the violation of Dashen's theorem at tree level due to those quark mass differences is significant numerically for kaons. In our present treatment
  we consider those differences on the same footing as the EM corrections to the masses. For pions, the main effect in the $\pi^0-\pi^+$ mass difference comes from the EM contribution \cite{Das:1967it}.

All the previous expressions hold for tree level LO masses $M_a^2$ with $a=\pi^{\pm},\pi^0,K^{\pm},\eta$, in terms of which we will  write all of our results.  They
 coincide with the physical masses
 to leading order in ChPT, i.e., $M_{a,phys}^2=M_a^2(1+\Od(M^2))$. Calculating the ChPT corrections to a given order allows then to determine the numerical values of the tree level masses,
 knowing their physical values and to that order of approximation.
 The same holds for $F$, which coincides with the meson decay constants in
 the chiral limit $F_{a,phys}^2=F^2(1+\Od(M^2))$. Next to leading order $\Od(M^2)$ corrections to meson masses
 and decay constants were given in \cite{Gasser:1983yg,Gasser:1984gg} for $e^2=0$. EM corrections to the masses
  can be found in \cite{Urech:1994hd} for $SU(3)$ and in \cite{Knecht:1997jw,Schweizer:2002ft} for $SU(2)$ including both $e^2\neq 0$ and
$m_u\neq m_d$ isospin-breaking terms.

The fourth-order lagrangian  in (\ref{efflag}) consists of all possible terms compatible with the QCD symmetries to that order, including the EM ones.  The ${\cal L}_{p^4}$ lagrangian is given in \cite{Gasser:1983yg} for the $SU(2)$ case, $h_{1,2,3}$
 (contact terms) and $l_{1\dots 7}$ denoting the dimensionless  LEC multiplying each independent term, and in \cite{Gasser:1984gg}
 for $SU(3)$  the LEC named  $H_{1,2}$ and $L_{1\dots 10}$. The EM ${\cal L}_{e^2p^2}$ and ${\cal L}_{e^4}$ for $SU(2)$
 are given in \cite{Meissner:1997fa,Knecht:1997jw}, $k_{1,\dots 13}$ denoting the corresponding EM LEC, and in \cite{Urech:1994hd}
 for $SU(3)$ with the $K_{1\dots17}$ EM LEC. For completeness, we give in Appendix \ref{app:lag} the relevant terms needed in this
 work. The LEC are renormalized in such a way that they absorb all the one-loop ultraviolet divergences coming
 from ${\cal L}_{p^2}$ and ${\cal L}_{e^2}$, according to the ChPT counting, and depend on the $\overline{MS}$ low-energy renormalization scale
 $\mu$ in such a way that the physical quantities are  finite and scale-independent. The renormalization conditions for all  the LEC can be
 found  in
 \cite{Gasser:1983yg,Urech:1994hd,Neufeld:1995mu,Knecht:1997jw} and we
 collect in Appendix \ref{app:lag} only those needed in the
present work.

As customary, we denote the  scale-dependent and renormalized
    LEC  by an ``$r$" superscript. The renormalized LEC are independent of the quark masses by definition, although their finite parts are unknown, i.e., they are not provided within the low-energy theory. The numerical values of the LEC at
    a given scale can be estimated by fitting  meson experimental data, theoretically by matching
    the underlying theory under some approximations, or from the lattice. These procedures allow to obtain estimates
    for the ``real-world" LEC at the expense of introducing residual
dependencies of those LEC on the parameters of the approximation
procedure, which typically involves a truncation of some kind.
Examples of these are the $m_s$ dependence on the $SU(2)$ LEC when
matching the $SU(3)$ ones, the correlations between LEC, masses
and decay constants through the fitting procedure, the QCD
renormalization scale  and  gauge dependence of some of the EM LEC
or the dependence with  lattice artifacts such as finite size or
spurious meson masses. We will give more details below, specially
regarding the EM LEC which will play an important role in our
present work. An exception to the LEC estimates are the contact
LEC $h_i$ and $H_i$, which are needed for renormalization but
cannot
    be directly measured. The  physical quantities depending on them are  therefore ambiguous, which comes from the  definition of the condensates in QCD perturbation theory, requiring subtractions to converge
    \cite{Gasser:1983yg}. It is therefore phenomenologically convenient to define suitable combinations which are independent of
    the $h_i,H_i$. We will bear this in mind throughout this work, providing such combinations when isospin-breaking is included.

We will analyze in one-loop ChPT (next to leading order)  the quark condensates, which for a given flavour $q_i$ can be written at that order as:

\begin{eqnarray}
 \condtwoi= -\left\langle \frac{\partial {\cal L}_{eff}}{\partial m_i}\right\rangle.
\label{conddef}
\end{eqnarray}

The above equation is nothing but the functional derivative with
respect to the $i$-th component of the scalar current,
particularized to the values of the physical quark masses,
according to the external source method \cite{Gasser:1983yg,Gasser:1984gg}.
Therefore, we will be interested only in the terms of the
fourth-order lagrangian containing at least one power of the quark
masses. These are the operators given in eqns.(\ref{lag4su2}) and
(\ref{lag4su3}) for  $SU(2)$ and
$SU(3)$  respectively. Thus, the  LEC that enter in our
calculation are $l_3,h_1,h_3,k_5,k_6,k_7$ in $SU(2)$, and
$L_6,L_8,H_2$, $K_7,K_8,K_9,K_{10}$ in $SU(3)$. Besides, up to
NLO, only tree level diagrams from the fourth-order lagrangian can
contribute to the condensates, so that in practice it is enough to
set $U=\mathds{1}$ in (\ref{lag4su2})-(\ref{lag4su3}) for getting
those tree-level contributions from (\ref{conddef}).

\subsection{Masses and  low-energy constants}
\label{sec:numval}

For most of the numerical values of the different low-energy constants and parameters in the $SU(3)$ case, we will follow \cite{Amoros:2001cp}, where fits
 to $K_{l4}$ experimental data are performed in terms of $\Od(p^6)$ ChPT expressions,  including the isospin mass difference $m_u/m_d\neq 1$ and EM  corrections to the meson masses, extending a previous work \cite{Amoros:2000mc} where isospin breaking was not considered.  Those fits have been improved in a recent work \cite{Bijnens:2011tb}, which takes into account new phenomenological and lattice results. We will however stick to the values of \cite{Amoros:2001cp}, since our main interest is to compare the isospin-breaking condensates with the two sources included and to estimate the effect of the  EM LEC. In the new fits \cite{Bijnens:2011tb} isospin breaking is included only to correct for the charged kaon mass and  the condensate values are not provided. For a review of different estimates of the quark masses and condensates see also \cite{Narison:2002hk} and \cite{Ioffe:2005ym}. In addition, in \cite{Colangelo:2010et} a recent
 update  of lattice results for low-energy parameters can be found, including LEC and the quark condensate. We will use the central values of the main fit in \cite{Amoros:2001cp}. The value of $m_s/\hat m=24$ \cite{Narison:2002hk,Ioffe:2005ym}  is used as an input in \cite{Amoros:2001cp}, as well as $L_6^r=0$, as follows e.g. from OZI rule or large-$N_c$ arguments \cite{Gasser:1984gg}.  The more recent fits \cite{Bijnens:2011tb} consider an updated value of  $m_s/\hat m=27.8$ in accordance with recent determinations \cite{Colangelo:2010et} and a nonzero value of $L_6^r$ is obtained as an output. The
 suppression of $L_6^r$ has been questioned  in connection with a reduction of the light quark condensate when the number of flavours is increased \cite{sternmouss}, within the framework of generalized ChPT. In that context, the chiral power counting is modified due to the smallness of the condensate.
  Here, we will adhere to the standard ChPT picture, where the condensate and the GOR-like relations are dominated by the leading order \cite{Colangelo:2001sp}, sustained by the  recent lattice LEC estimates \cite{Colangelo:2010et}.  The values of $F=87.1$ MeV, $2B_0\hat m=0.0136$ GeV$^2$, $m_u/m_d=0.46$ and $L_8(\mu=770$ MeV$)=0.62\times 10^{-3}$ are outputs from the main fit in \cite{Amoros:2001cp}. With those values we get from (\ref{mixangle}) $\varepsilon=0.014$ and from (\ref{treemassessu3}) the tree level masses of $\pi^0,K^0,\eta$.

  To calculate the tree level  charged meson masses, we  need also the value of the $C$ constant, which can be inferred also from the results in \cite{Amoros:2001cp} since the EM correction is numerically very small in  the charged kaon mass  with respect to the pure QCD contribution. This allows to extract the tree level charged kaon mass directly from the  expressions for $M_{K^\pm}/M_{K^\pm,QCD}$ in \cite{Amoros:2001cp}, approximating $M_{K^\pm,QCD}$ by the full physical mass. From there we extract the value of $C$ by subtracting the tree level QCD part in (\ref{treemassessu3}) calculated with the above given quark masses. In this way we get $C=5.84\times 10^7$ MeV$^4$, which is very close to the values obtained simply from the charged-neutral pion mass difference in (\ref{treemassessu3}) setting the masses and $F$ to their physical values \cite{Knecht:1997jw} or from resonance saturation arguments \cite{Urech:1994hd}.
  From that $C$ value we obtain the tree-level charged pion mass,  using again (\ref{treemassessu3}). Nevertheless, to the order we are calculating we could have used as well the physical meson masses and decay constants instead of the tree level ones, since formally the difference is hidden in higher orders.  The main reason why we choose the values in \cite{Amoros:2001cp} is to compare directly with their numerical quark condensates and estimate the importance of the $K_i^r$ corrections (see section \ref{sec:resultsu3} for details). The constant $H_2$ will also appear explicitly in quark condensates. Since it cannot be fixed with meson experimental data, when needed we will estimate it from scalar resonance saturation arguments as $H_2^r=2L_8^r$ \cite{Ecker:1988te,Amoros:2001cp}, although we will comment below more about the $H_2^r$ dependence of the results and provide
  physical quantities which are independent of the contact terms.

Regarding the EM LEC, the $SU(3)$  $K_i^r$  have been estimated in
the literature under different theoretical schemes. Resonance
saturation was used in \cite{Baur:1996ya}, large-$N_c$ and NJL
models in \cite{Bijnens:1996kk}, complemented with QCD
perturbative information in \cite{Pinzke:2004be} and a sum-rule
approach combined with low-lying resonance saturation has been
followed in \cite{Moussallam:1997xx,Ananthanarayan:2004qk}. The
works
\cite{Bijnens:1996kk,Pinzke:2004be,Moussallam:1997xx,Ananthanarayan:2004qk}
have in common the use of perturbative QCD methods for the
short-distance part of the LEC and  different model approaches for
the long-distance part. This procedure implies that the LEC
estimated in that way depend (roughly logarithmically) in general
on the QCD renormalization scale,  which we call $\mu_0$ to
distinguish it from the low-energy scale $\mu$, as well as on the
gauge parameter. A closely related problem is that the separation
of the strong ($e=0$) and EM contributions for a
given physical quantity is in principle ambiguous
\cite{Bijnens:1993ae,Bijnens:1996kk,Moussallam:1997xx,Gasser:2003hk}.
 The origin of this
ambiguity \cite{Gasser:2003hk} is that QCD scaling quantities such as quark masses
contain also EM contributions through the Renormalization Group
(RG) evolution in the full QCD+EM theory. Thus, a particular
prescription for disentangling those contributions must be
provided.  In addition, when matching such
quantities between the low-energy sector and the underlying
theory, the choice of a given prescription will necessarily affect
the scale and gauge dependence of the EM LEC. These theoretical
uncertainties, as well numerical ones, make those
theoretical EM LEC estimates not fully compatible among them.  For
these reasons, in many works analyzing  EM corrections, the
LEC are simply assumed to lie within ``natural" values $\vert
K_i^r \vert$, $\vert k_i^r \vert$ $\lsim \frac{1}{16\pi^2}\simeq
6.3\times 10^{-3}$ at the scale $\mu\sim M_\rho$
\cite{Urech:1994hd,Knecht:1997jw}. The above theoretical issues  will be addressed in
more detail  in sections \ref{sec:condsu2} and
\ref{sec:boundssu3}. The LEC dependencies on the QCD scale and on
the gauge parameter do no affect directly our results, being only
relevant when comparing them with approaches where those LEC
are obtained by matching the underlying theory. In
that context we will see that the LEC combinations that we will
deal with are gauge independent and lie within the  stability range where the dependence on $\mu_0$ is smooth and the matching makes sense \cite{Bijnens:1996kk,Pinzke:2004be}. The theoretical errors quoted e.g. in
\cite{Bijnens:1996kk} account for the uncertainty related to the $\mu_0$ dependence.

 As for the $SU(2)$ $k_i^r$, no direct estimate is available to our knowledge, although one can relate them to the $K_i^r$ by performing formally a $1/m_s$ expansion in a given
 physical quantity calculated in $SU(3)$ and comparing to the corresponding $SU(2)$ expression,
similarly as the $l_i\leftrightarrow L_i$ conversion given in
\cite{Gasser:1984gg}. This has been done partially for some
combinations of the LEC, namely, those appearing in the neutral pion
mass  \cite{Gasser:2001un}, in pion scattering
\cite{Gasser:2001un,Knecht:2002gz} and in the pionium lifetime
\cite{Gasser:2001un,Jallouli:1997ux}. More recently, a full
matching of the EM $SU(2)$ and $SU(3)$ LEC at the lagrangian level
has been performed in \cite{Haefeli:2007ey} using functional
integral methods  in the chiral limit. In the present work, we
will provide a complementary analysis. Namely, in section
\ref{sec:condsu3} we will obtain the matching relations between
the LEC involved in the quark condensates, including both
$m_u-m_d$ and EM contributions. Those relations will be  consistent with the results in \cite{Haefeli:2007ey}
 and phenomenologically useful when dealing with
 approximate LEC determinations  where isospin
and mass corrections may be entangled.

\section{Two-flavour quark condensates and bounds for the EM LEC}
\label{sec:condsu2}

  We start by giving the explicit one-loop ChPT expressions for the quark condensates
in $SU(2)$ with all the isospin-breaking corrections included,
which we derive from (\ref{conddef}):

\begin{widetext}
  \begin{eqnarray}
  \condtwo\equiv \condsum&=&-2F^2B_0\left[1-\mu_{\pi^0}-2\mu_{\pi^\pm}+2\frac{M_{\pi^0}^2}{F^2}\left(l_3^r(\mu)+h_1^r(\mu)\right) +e^2\mathcal{K}_2^r(\mu)+\mathcal{O}\left(p^4\right)\right],
\label{condsu2sum}\\  \conddif&=&4B_0^2(m_d-m_u)h_3-\frac{8}{3}F^2B_0e^2k_7 + \mathcal{O}\left(p^2\right),
\label{condsu2dif}
  \end{eqnarray}
  \end{widetext}
  where:

   \begin{equation}
   \mathcal{K}_2^r(\mu)=\frac{4}{9}\left[5\left(k_5^r(\mu)+k_6^r(\mu)\right)+k_7\right],
   \label{K2def}
   \end{equation}
 and   throughout this work we will follow the same  notation as in \cite{Gasser:1983yg,Gasser:1984gg}:

   \begin{equation}
   \mu_i=\frac{M_i^2}{32\pi^2 F^2}\log\frac{M_i^2}{\mu^2} \quad ; \quad \nu_i=\frac{1}{32\pi^2}\left[1+\log\frac{M_i^2}{\mu^2}\right].
   \label{munudef}
   \end{equation}

   The $\mu_i$ arise  from the finite part of the one-loop tadpole-like contribution $G_i(x=0)$, with $G$ the free meson propagator
   \cite{Gasser:1983yg}. The renormalization conditions for the LEC
involved in (\ref{condsu2sum})-(\ref{condsu2dif}) can be found  in
\cite{Gasser:1983yg} and \cite{Knecht:1997jw} and we give them in
Appendix \ref{app:lag}.  With that LEC renormalization, one can
check that the condensates in
(\ref{condsu2sum})-(\ref{condsu2dif}) are finite and
scale-independent, which is a nontrivial consistency check. Recall
that $h_3$ and $k_7$ do not need to  be  renormalized.
 The condensates still depend on the $h_1$ and $h_3$ contact LEC,
which, as explained above, yield an ambiguity in the determination of the condensates. The result (\ref{condsu2sum}) for $e\neq 0$ and $m_u\neq m_d$
reduces for $e=0$  to the expressions given in
\cite{Gasser:1983yg}. The condensate difference
(\ref{condsu2dif}) is given in \cite{Gasser:1983yg} for $e=0$,
which we agree with, and in \cite{Knecht:1997jw} for $e\neq 0$,
which we also agree with, except for the relative sign between the
two terms, which should be a minus in their eq.(3.12)
\cite{knechtpriv}.

Let us now argue on how the EM corrections to the condensates may lead to constraints for the EM LEC. Those corrections  come directly from the coupling of the EM field to the quarks in the QCD action, which break chiral symmetry. Actually, to understand better the origin of the different sources involved in chiral symmetry and isospin breaking, it is useful to keep the charges of the $u$ and $d$ flavours arbitrary and to  separate the isoscalar and isovector contributions of the charge matrix in $SU(2)$:

\begin{equation}
Q=\frac{e_u+e_d}{2}\mathds{1}+\frac{e_u-e_d}{2}\tau_3,
\label{chargesep}
\end{equation}
with  $\tau_3=\diag(1,-1)$ corresponding to the third isospin
component.  The EM part of the QCD lagrangian $\bar q \gamma^\mu
A_\mu Q q$ breaks explicitly
chiral symmetry  $SU_L(2)\times SU_R(2)$  if $e_u\neq
e_d$,  through the isovector part in (\ref{chargesep}). The
isovector  also breaks the isospin symmetry $SU_V(2)$ ($L=R$)
except for transformations in the third direction, which
corresponds to electric charge conservation. On the other hand,
the mass term $\bar q {\cal M} q$ breaks chiral symmetry for any
nonzero value of the quark masses, preserving isospin symmetry if
$m_u=m_d$. Altogether, the conclusion is that the QCD lagrangian
is chiral invariant only if  $e_u=e_d$ {\em and} $m_u=m_d=0$.
Thus,   chiral symmetry is explicitly broken even
if $e_u=e_d$, as long as any of the quark masses $m_q\neq 0$, or
equivalently, in the presence of an external scalar source, as
needed to derive the condensates. If $e_u=e_d$ and $m_u=m_d\neq
0$, chiral symmetry is broken but isospin symmetry is conserved.

Now, let us remind  how this charge and mass symmetry breaking pattern translates into the low-energy sector. The leading order ${\cal L}_{p^2+e^2}$ in (\ref{L2}) contains
 separate combinations of the charge and mass terms,  both sharing the QCD pattern. Thus,  the charge contribution proportional to $C$ in (\ref{L2}) can be decomposed according to (\ref{chargesep}), giving a constant term proportional to $(e_u+e_d)^2$ independent of masses and fields, plus the term $C\left[(e_u-e_d)^2/4\right]\tr\left[\tau_3 U \tau_3 U^\dagger\right]$, which contributes  directly to the pion EM mass difference in (\ref{treemassessu2}). Therefore, in the second order lagrangian all the EM chiral symmetry breaking terms are proportional to $(e_u-e_d)^2$. This is no longer true for the fourth order lagrangian in (\ref{lag4su2}), for which the symmetries of the theory allow for crossed mass-charge terms, like those proportional to $k_5,k_6$ and $k_7$.  Those crossed terms break chiral symmetry even for $e_u=e_d$ for any
 nonzero quark mass, the strength of chiral breaking being proportional both to the quark charge $e$ and to the quark mass $\hat m$. Consequently, they contribute to $\condtwo=-2\langle\partial{\cal L}/\partial(m_u+m_d)\rangle$, the expectation value of the $SU_V(2)$ singlet behaving as an order parameter for chiral symmetry breaking. Setting $U=\mathds{1}$ in the lagrangian gives a piece proportional to $k_5+k_6$ yielding   $(e_u+e_d)^2$ and
$(e_u-e_d)^2$ contributions to $\condtwo$ and another one proportional to $k_7$ contributing both to $\condtwo$ and  to $\conddif=-2\langle\partial{\cal L}/\partial(m_u-m_d)\rangle$, the isotriplet order parameter of isospin breaking:

 \begin{eqnarray*}
 {\cal L}_{e^2p^2}^{\condtwo}&=&2F^2 B_0\left\{(k_5+k_6)\left[(e_u+e_d)^2+(e_u-e_d)^2\right]
(m_u+m_d)\right.\nonumber\\&+&\left.2k_7\left[(e_u+e_d)^2(m_u+m_d)+(e_u-e_d)(e_u+e_d)(m_u-m_d)\right]\right\}+\dots
 \end{eqnarray*}
 where the dots indicate terms not contributing to the
condensates at this order.

 Thus,  we see how the two
sources of isospin breaking show up in the order parameter (\ref{condsu2dif}), which does not receive pion
loop contributions in $SU(2)$. The latter is the explicit confirmation that isospin symmetry is not spontaneously broken in QCD \cite{Vafa:1983tf}, since all the contributions to this order parameter vanish for $m_u=m_d$ and $e=0$.

On the other hand, the quark condensate $\condtwo$ behaves as an order parameter for  chiral symmetry and therefore measures the different sources of symmetry breaking: spontaneous and explicit. Thus, it is naturally expected that its absolute value increases when a new symmetry-breaking source, as the EM contribution,  is switched on. This increasing behaviour is what we denote as ferromagnetic, in analogy with the behaviour of the magnetization in a ferromagnetic material under an external magnetic field.

There is no a priori formal argument to ensure the ferromagnetic-like nature of the QCD low-energy vacuum. We can nevertheless learn from the response of the system to the light quark mass $\hat m$, which is the actual counterpart of the magnetic field in a ferromagnet, since it breaks the chiral symmetry explicitly by coupling to the order parameter.   For  the mass, this ferromagnetic behaviour  is actually  followed by the  $e=0$ condensates in the standard ChPT framework, both to $\Od(p^4)$ and to $\Od(p^6)$ with the LEC in \cite{Amoros:2000mc}, although assuming  $L_6$ suppression and the dependence on contact terms still introducing a source of ambiguity. With the recent fit giving nonzero $L_6^r$ and a new value also for $L_8^r$ \cite{Bijnens:2011tb} we get  also that $\condsum$ increases  with light quark masses, from our $\Od(p^4)$ expressions.  The same ferromagnetic effect  implies the increasing of the critical temperature of chiral restoration when increasing the pion mass, confirmed by ChPT calculations \cite{Gerber:1988tt} and lattice simulations \cite{Cheng:2009zi}. Finally, lattice results for the condensate \cite{Colangelo:2010et} reveal a systematic increase of its absolute value with respect to direct estimates \cite{Narison:2002hk}, reflecting again the same behaviour, since the pion masses used in the lattice remain above the physical values. The EM symmetry-breaking is of different nature as the mass,  the former coming from vector-like interactions while the later being of scalar type. However, as we have discussed in the previous paragraphs,  their symmetry breaking effects on certain observables are similar. Thus, the isovector part in (\ref{chargesep}) increases the masses of the charged mesons, according to  (\ref{treemassessu2}) and (\ref{treemassessu3}),  while the isovector and isoscalar both mix with the mass and contribute to $\condtwo$. There are other arguments pointing in the same direction when EM interactions are switched on. At finite temperature, the EM pure thermal corrections to the condensate also increase its absolute value for any temperature \cite{Nicola:2011gq}. On the other hand, the condensate increases under the influence of an external magnetic field $eH$, which can be also  understood as the reduction of the free energy $\epsilon\sim m_q\condtwo<0$ (to leading order) needed to compensate for the EM energy increasing $\Delta\epsilon_{EM}\sim (eH)^2/2>0$ \cite{Shushpanov:1997sf}.

Our purpose here will be to explore the consequences of that EM ferromagnetic behaviour to leading order. If the vacuum response is ferromagnetic, certain bounds for the EM LEC involved should be satisfied. We will derive those bounds and show that they are independent of the low-energy scale and thus can be checked in terms of physical quantities. Next we will check that the bounds are satisfied for the different estimates available for the EM LEC, with more detail for  the $SU(3)$ case in section \ref{sec:boundssu3}, where we also discuss the gauge independence of our results. This will provide a consistency check for the ferromagnetic behaviour.

An important comment is that we will discuss the ferromagnetic-like
condition on the EM correction to $\condtwo$ and, as explained
above, the splitting of the $e=0$ and $e\neq 0$ parts in QCD+EM is
ambiguous \cite{Bijnens:1993ae,Gasser:2003hk}. This does not affect the low-energy representation of the condensates, which can be written in terms of physical quantities such as meson masses and decay constants. However, we will test our bounds with the EM LEC estimates obtained by matching  low-energy results with the underlying theory \cite{Bijnens:1996kk,Baur:1996ya,Pinzke:2004be,Moussallam:1997xx}. Therefore, we have to be consistent with the  prescription for  charge splitting followed in those works. This amounts
to the direct separation  of the $e=0$ part, which still  may contain residual charge and $\mu_0$ QCD scale dependence through running parameters. The consequence is that the EM LEC
thus defined are in general $\mu_0$-dependent, as discussed in \cite{Gasser:2003hk}. Therefore, those estimates
are reliable only if there is a stability range  where
the dependence on $\mu_0$ is smooth and lies within the
theoretical errors \cite{Bijnens:1996kk}. Actually, such stability range criteria are met for the LEC involved in our analysis (see section \ref{sec:boundssu3}). Within that range, our identification of the $e^2$ dependent part in $\condtwo$ is consistent with the splitting scheme followed in those works. Actually, in that scheme $F_0$ is $\mu_0$-independent and the $e^2$-dependent part of the $\mu_0$ running of $B_0 m_{u,d}$ is the same as that of $m_{u,d}$ in perturbative QCD+EM \cite{Gasser:2003hk}. Hence, it is consistent to assume that the ChPT leading order of $\condtwo= -2B_0F_0^2+\dots$ does not introduce any residual $e^2$ dependence when performing the charge splitting in the low-energy expression. Using a different splitting prescription  would lead in general to different bounds and a different definition of the EM LEC. For instance, an alternative splitting procedure is  introduced in \cite{Gasser:2003hk} by matching running parameters of the $e^2$ theory with those of the $e^2=0$ one at a given matching scale $\mu_1$. In that way, the  EM part can be chosen as $\mu_0$ independent but it depends on the matching scale $\mu_1$. We will not consider that splitting here, since there are no available theoretical estimates for the LEC defined with that scheme. The  scale dependence for the LEC in either scheme is roughly expected to be logarithmic.

 Having the above considerations in mind, and going back to the case of physical quark charges, we separate the EM corrections to the condensate through  the ratio:

\begin{eqnarray}
\frac{\condtwo^{e\neq 0}}{\condtwo^{e=0}}=1+2\left[\mu_{\pi^0}-\mu_{\pi^{\pm}}\right]+e^2\mathcal{K}_2^r(\mu)+\mathcal{O}\left(p^4\right)
\nonumber\\
=1+e^2\mathcal{K}_2^r(\mu)-\frac{4Ce^2}{F^4}\nu_{\pi^0}+\Od(\delta_\pi^2)+\Od(p^4),
\label{emratiosu2}
\end{eqnarray}
with $\nu_i$ defined in (\ref{munudef}) and where we have expanded in $\delta_\pi\equiv\left(M_{\pi^\pm}^2-\!M_{\pi^0}^2\right)/M_{\pi^0}^2\simeq 0.1$, as $\mu_{\pi^{\pm}}-\mu_{\pi^0}=\delta_\pi M_{\pi^0}^2(\nu_{\pi^0}/F^2)+\Od(\delta_\pi^2)$, which
is numerically reliable and can be performed in addition to the chiral expansion, in order to simplify the  previous expression.

 We note that in $SU(2)$ and to this order, the ratio (\ref{emratiosu2})  is not only finite and scale-independent but it is also independent of the not-EM LEC, including the contact  $h_1,h_3$, and therefore free of ambiguities related to the condensate definition.   In fact, this ratio is also independent of $B_0$, unlike the individual quark condensates, which have only physical meaning and give rise to
 observables when multiplied by the appropriate quark masses, since  $m_i B_0\sim M_i^2$. In $SU(2)$, the above ratio does not depend either on the mass difference $m_d-m_u$, i.e., it depends only on the  sum $\hat m$ and its deviations from unity are therefore  purely of EM  origin. All these  properties make the ratio (\ref{emratiosu2})  a suitable quantity to
 isolate the EM effects on the condensate. Thus, the ferromagnetic-like nature of the chiral order parameter $\condtwo$, within its low-energy representation,  would require that this ratio is greater than one, or equivalently to this order, $\partial\condtwo/\partial e^2\geq 1$. That condition leads to the following lower bound for the combination of EM LEC involved to this order, neglecting the $\Od(\delta_\pi^2)$ in (\ref{emratiosu2}) which changes very little the numerical results:

\begin{equation}
5\left[k_5^r(\mu)+k_6^r(\mu)\right]+k_7^r\geq \frac{9C}{F^4}\nu_{\pi^0}.
\label{su2bound}
\end{equation}

 We remark  that the  bound (\ref{su2bound})  is  independent of the low-energy
scale $\mu$ at which the LEC on the left-hand side are evaluated as long as the same scale is used on
the right-hand side. Thus, it provides a well-defined low-energy prediction, expressed in terms of meson masses. The LEC on the left-hand side could be estimated by fitting low-energy processes or theoretically from the underlying theory, with all the related subtleties commented above.

The condition (\ref{su2bound}) and the corresponding ones for $SU(3)$ that will be derived in section \ref{sec:boundssu3} are obtained as a necessary condition that the LEC should satisfy if the QCD physical vacuum is ferromagnetic. This positivity condition on the quark condensate probes the vacuum by taking the mass derivatives (\ref{conddef}) through the external source method so that the quark masses have to be kept different from zero  and in that way  the explicit symmetry breaking corrections are revealed in the condensate.  If one is interested in the chiral limit, it must be taken only {\em after} differentiation, i.e., directly in eq.(\ref{emratiosu2}). In that case, it is not justified to perform the additional $\delta_\pi$ expansion in the charge because the $e=0$ masses vanish and one would be left only with the $\mu_{\pi^\pm}$ contribution in the r.h.s. of (\ref{emratiosu2}), now with $M^2_{\pi^\pm}=2Ce^2/F^2$. That would actually give a larger negative value for the lower bound, coming from the smallness of the charged part of the pion mass (see details below) so that the bound in the chiral limit is less predictive.  The fact that  our bounds depend on quark masses is similar to other bounds on LEC obtained from QCD inequalities \cite{Comellas:1995hq}.

 Nevertheless, the main physical interest is to test this bound for physical masses, using  different estimates of the LEC in the literature. Thus, as a rough estimate, setting $\mu=M_\rho\simeq$ 770 MeV and with  the physical pion masses $M_{\pi^\pm}\simeq 139.57$ MeV,
$M_{\pi^0}\simeq 134.97$ MeV,  the  bound (\ref{su2bound})
gives $5\left[k_5^r(M_\rho)+k_6^r(M_\rho)\right]+k_7^r\geq -6.32
(5.62)\times 10^{-2}$ taking $F=87.1 (92.4)$ MeV. This is a bit
more restrictive than the ``natural" lower bound $-6.93\times
10^{-2}$ for the above LEC combination, obtained by setting all of
them to $-1/(16\pi^2)$. The chiral limit gives $-0.17$ (with the  value of $C$ discussed in section \ref{sec:numval} and $F=87.1$ MeV) i.e., much less restrictive, as commented above.  More detailed numerical analysis  will be done for $SU(3)$ in section \ref{sec:boundssu3}.

On the other hand, the maximum value for
the ratio (\ref{emratiosu2}) for the $k_i^r$ within ``natural" values is
obtained by setting the three of them to
$k_i^r(\mu=M_\rho)=1/(16\pi^2)$, giving $\frac{\condtwo^{e\neq 0}}{\condtwo^{e=0}}=1.0054$, which
gives an idea of the size of this  correction. We  remark that the term proportional to $\nu_\pi$ on the ratio (\ref{emratiosu2}) comes directly from the dependence of the pion masses on $e^2$, so that it parametrizes the corrections in the condensate coming from any source of pion mass increasing, not only the EM one. Therefore, the same result
 can be used in order to provide a rough estimate of lattice errors in the condensate due to including heavier pion masses as lattice artifacts.  In some lattice algorithms like the staggered fermion one, the situation is very similar to the mass differences induced by the charge terms. In that formalism,  the finite
   lattice spacing induces terms \cite{Aubin:2003mg} that break explicitly the so called taste symmetry (four different quark species or ''tastes" are introduced for every quark flavour) leaving a residual $U(1)$ symmetry, pretty much in the same way as the charge term in (\ref{L2}).  As a rough estimate, we can then replace   $2Ce^2/F^4$ by the corresponding $\delta_\pi$ from the lattice, obtained as the difference between the mass of the lightest lattice meson and the true pion mass. For a lattice pion mass of about 300 MeV, the $\nu_\pi$ term in (\ref{emratiosu2}) gives a correction of about $6\%$, which for a condensate value of (250 MeV)$^3$ represents about (5 MeV)$^3$, which is within  the order of magnitude  quoted in  \cite{Colangelo:2010et}. Nevertheless, it should be taken into account that the staggered ChPT \cite{Aubin:2003mg} has a much richer structure than the EM terms considered here and in particular there will be other operators contributing to the condensates at tree level, multiplied by the pertinent low-energy constants. If those constants are of natural size, we expect the size of the  corrections to the condensate to remain within the range  quoted above.

  \section{Three flavour quark condensates.}
 \label{sec:condsu3}
\subsection{Results for light and strange condensates.}
\label{sec:resultsu3}

In the $SU(3)$ case, we  derive to one loop in ChPT the light condensate $\condtwo_l=\langle \bar u u + \bar d d\rangle$  and the strange one $\langle \bar s s \rangle$, taking into account both $m_u-m_d$ and $e\neq 0$ corrections. Apart from the kaon and eta loops, an important  distinctive feature in this case is the appearance of the $\pi^0\eta$ mixing term with the tree-level mixing angle $\varepsilon$ defined in (\ref{mixangle}) which is one of the sources of isospin-breaking corrections. The results we obtain for the condensates with all the corrections included are the following:

\begin{widetext}
\begin{eqnarray}
  \condtwo_l^{SU(3)}\equiv \condsum^{SU(3)}&=&-2F^2B_0\left\{1+\frac{8B_0}{F^2}\left[\hat m\left(2L_8^r(\mu)+H_2^r(\mu)\right)+4(2\hat m + m_s)L_6^r(\mu)\right]+e^2\mathcal{K}_{3+}^r (\mu)\right.\nonumber\\&-&\left.\frac{1}{3}\left(3-\sin^2 \varepsilon\right) \mu_{\pi^0}-2\mu_{\pi^\pm}-\mu_{K^0}-\mu_{K^\pm}-\frac{1}{3}\left(1+\sin^2 \varepsilon\right)\mu_\eta  +\mathcal{O}\left(p^4\right)\right\}
\label{condsu3sum}\\
\conddif^{SU(3)}&=&2F^2B_0\left\{\frac{4B_0}{F^2}(m_d-m_u)\left(2L_8^r(\mu)+H_2^r(\mu)\right)-e^2\mathcal{K}_{3-}^r (\mu)
\right.\nonumber\\&+&\left. \frac{\sin 2\varepsilon}{\sqrt{3}}\left[\mu_{\pi^0}-\mu_{\eta}\right]+\mu_{K^\pm}-\mu_{K^0}\right\}+\mathcal{O}\left(p^2\right)
\label{condsu3dif}\\
\langle \bar s s \rangle&=&-F^2B_0\left\{1+\frac{8B_0}{F^2}\left[m_s\left(2L_8^r(\mu)+H_2^r(\mu)\right)+4(2\hat m + m_s)L_6^r(\mu)\right]+e^2\mathcal{K}_{s}^r (\mu)
\right.\nonumber\\&-&\left.
\frac{4}{3}\left[ \mu_{\pi^0}\sin^2 \varepsilon+\mu_{\eta}\cos^2 \varepsilon\right]-2\left[\mu_{K^\pm}+\mu_{K^0}\right]+\mathcal{O}\left(p^4\right)\right\}
  \label{condsu3str}\end{eqnarray}
  \end{widetext}
where we use the notation (\ref{munudef}) and:
\begin{eqnarray}
\mathcal{K}_{3+}^r(\mu)&=&\frac{4}{9}\left[6\left(K_7+K_8^r(\mu)\right)+5\left(K_9^r(\mu)+K_{10}^r(\mu)\right)\right],\nonumber\\
\mathcal{K}_{3-}^r(\mu)&=&\frac{4}{3}\left[K_9^r(\mu)+K_{10}^r(\mu)\right],\nonumber\\
\mathcal{K}_{s}^r(\mu)&=&\frac{8}{9}\left[3\left(K_7+K_8^r(\mu)\right)+K_9^r(\mu)+K_{10}^r(\mu)\right].
\label{Kscondsu3}
\end{eqnarray}

Note that in some of the above terms we have preferred, for simplicity,  to leave the results in terms of quark  instead of meson masses. An important difference between the $SU(2)$ and $SU(3)$ cases is that now there are loop corrections in $\conddif$, where eta and pion loops enter through the mixing angle and kaon ones through the charged-neutral kaon mass difference. We have
checked  that the results are finite and scale-independent with the renormalization of the LEC given in Appendix \ref{app:lag} and that they  agree with \cite{Gasser:1984gg} for $e=0$. Some unpublished results related to the $SU(2)$ and $SU(3)$ isospin-breaking condensates  can also be found in \cite{Nehme}.

Numerical results for the
condensates to this order can be found in \cite{Amoros:2001cp}. As
explained in section \ref{sec:numval}, the effect of the $K_i^r$
constants (\ref{Kscondsu3}) in the condensates is not fully considered in that work,  where the EM contributions are included
 through the corrections of Dashen's theorem \cite{Bijnens:1996kk}, so that only the $K_i^r$ combinations entering mass
renormalization appear. We will use then our results with all corrections included  to estimate the range of sensitivity to the $K_i^r$ of the condensates,  analyzing the possible relevance for the  fit in \cite{Amoros:2001cp}. Our results are displayed in Table \ref{tab:condval}. As discussed in section \ref{sec:numval}, we take the same input values $L_6^r=0$, $m_s/\hat m =24$ as in \cite{Amoros:2001cp} as well as the assumption $H_2^r=2L_8^r$, and the output values of  $B_0m_{u,d,s},m_u/m_d,F,L_8^r$ from their  main fit. In the second and third columns of Table \ref{tab:condval}, we give the results with all the EM $K_i^r$ fixed to their minimum and maximum ``natural" values. Since the  $K_i^r$ appear all with positive sign in  (\ref{Kscondsu3}),  the absolute values of the condensates obtained in this way are, respectively,  lower and upper bounds within the natural range. We compare with the results quoted in \cite{Amoros:2001cp} to the same $\Od(p^4)$ order (fourth column) for their main fit and we also show for comparison the results in the isospin limit $e=0$, $m_u=m_d$ (fifth column).  Our results agree reasonably with \cite{Amoros:2001cp}, although we note that the values in that work lie outside the natural range for the individual condensates. The largest relative corrections are about 2\% for the light condensates and about
  4\% for the strange one. These isospin-breaking corrections are significant given the precision of the condensates quoted in \cite{Amoros:2001cp}. On the other hand, the corrections  lie within the error range quoted by lattice analysis \cite{Colangelo:2010et}. In turn, note the bad ChPT convergence properties of the strange condensate, somehow expected since $\conds$ is much more sensitive to the strange quark mass $m_s$ than the light condensate \cite{Amoros:2000mc} and therefore the large strange explicit chiral symmetry breaking $m_s$ is responsible in this case for the spoiling of the ChPT series, based on perturbative mass corrections.  For the vacuum asymmetry $\frac{\condd}{\condu}-1$, the natural values band cover the result in \cite{Amoros:2001cp}, although the numerical discrepancies in that case are relatively larger,
  between 15\% and 24\% for the lower and upper limit of the $K_i^r$ respectively. Recall that this quantity vanishes to leading order in ChPT, according to (\ref{condsu3dif}), so that
   we expect it to be more sensitive to the $K_i^r$ correction, which in this case comes mostly  from the combination $K_9^r+K_{10}^r$. Nevertheless, it is worth noting that the results \cite{Amoros:2001cp} imply $\condd/\condu>1$ and $\conds/\condu>1$, both in disagreement with many sum rule estimates of the  condensate ratios \cite{Narison:2002hk}. Not surprisingly, we have the same discrepancy, since we use the same ChPT approach and the same numerical constants, except for the $K^r_i$ corrections. The discrepancy in the relatively large value of  $\conds/\condu$ comes possibly from the bad convergence of the ChPT series for the strange condensate, which in addition is very sensitive to the choice of $H_2^r$ \cite{Amoros:2001cp}. The light condensates converge much better and although the sign of $\condd/\condu-1$ is under debate, its magnitude is very small. In the latter case, our present calculation may become useful since the $K_9^r+K_{10}^r$ contribution may change the sign of the vacuum asymmetry, although its precise value to fit a given prediction for $\condd/\condu-1$ would still be subject to the $H_2^r$ value. For this reason, it is important to make predictions for quantities which are independent of this ambiguity, as we have done in section \ref{sec:condsu2} and as we will do in section  \ref{sec:sumrule}, where the sum rule for  condensate ratios will allow to make a more reliable estimate of  the vacuum asymmetry including both sources of isospin breaking. Finally,  we comment on the numerical differences by considering the more recent low-energy fits in \cite{Bijnens:2011tb}.  Still keeping  $H_2^r=2L_8^r$, these new values for the low-energy parameters  increase  considerably the total and strange condensates, which to $\Od(p^4)$ give $\condtwo/(2B_0F^2)\simeq -2.15$ and $\langle \bar s s \rangle/(B_0F^2)=-2.79$. These higher values are mostly due to the much smaller $F=65$ MeV,  obtained in the main fit of \cite{Bijnens:2011tb} to accommodate a rather high $L_4^r$ also with a large error $L_4^r=(0.75\pm 0.75)\times 10^{-3}$ (an output result in \cite{Bijnens:2011tb}). With the previous value $F=87.1$ MeV but keeping the rest of LEC and masses as in \cite{Bijnens:2011tb} we get $\condtwo/(2B_0F^2)\simeq -1.63$ and $\langle \bar s s \rangle/(B_0F^2)=-1.99$. The EM corrections remain of the same size and therefore their relative effect is somewhat smaller. As commented before,  $m_u\neq m_d$ isospin breaking is not implemented in those new fits and EM corrections are included only in kaon masses.

\begin{center}
\begin{table*}[ht]
{\small
\hfill{}
\begin{tabular}{|c|c|c|c|c|}
\cline{1-5}
\hline
  & $K^r_{7-10}=-\frac{1}{16\pi^2}$ & $K^r_{7-10}=\frac{1}{16\pi^2}$ & \cite{Amoros:2001cp} value $\Od(p^4)$ & Isospin limit   \\  \hline
 $-\condu_0/(B_0 F^2)$& 1.278 & 1.292 & 1.271 & 1.290 \\
\hline $-\condd_0/(B_0 F^2)$& 1.297 & 1.305 & 1.284 & 1.290 \\
\hline $-\conds_0/(B_0 F^2)$& 1.899 & 1.907 & 1.964 & 1.904\\
\hline $\frac{\condd}{\condu}-1$& 0.015 &0.010  & 0.013 & 0\\
\hline
\end{tabular}}
\hfill{}
\caption{\rm \label{tab:condval} Results for quark condensates. We compare with the values of \cite{Amoros:2001cp} to $\Od(p^4)$ using the same set of low-energy parameters as in the main fit of that work, except the $K_i^r$, which we consider at their lower (second column) and upper (third column) ``natural" values. We also quote the values in the isospin limit to the same chiral order.}
\end{table*}
\end{center}

\subsection{Sum rule corrections}
\label{sec:sumrule}

As noted in \cite{Gasser:1984gg}, for $m_u\neq m_d$ one can combine the isospin-breaking condensates into a sum rule relating the isospin asymmetry $\condd/\condu$ with the strange one $\conds/\condu$. Such relation is phenomenologically interesting because it does not include contact terms and hence  is suitable for  numerical estimates on the size of the isospin-breaking corrections. Our purpose in this section is to discuss the EM $e\neq 0$ contribution to that sum rule. To leading order in $m_u-m_d$ and $e^2$ we find:

\begin{widetext}
\begin{eqnarray}
  \Delta_{SR}\equiv \frac{\condd}{\condu}-1+\frac{m_d-m_u}{m_s-\hat m}\left[1-\frac{\conds}{\condu}\right]&=&\frac{m_u-m_d}{m_s-\hat m}\frac{1}{16\pi^2 F^2}\left[M_K^2-M_\pi^2+M_\pi^2\log\frac{M_\pi^2}{M_K^2}\right]\nonumber\\
  &+&e^2\left[\frac{C}{8\pi^2F^4} \left(1+\log\frac{M_K^2}{\mu^2}\right)-\frac{8}{3}\left(K_9^r(\mu)+K_{10}^r(\mu)\right) \right].
\label{sumrule0}
\end{eqnarray}
\end{widetext}

The last term, proportional to $e^2$  is scale-independent and is the charge correction to the result in \cite{Gasser:1984gg}. With the numerical set we have been using, the $m_d-m_u$ term in the right hand side gives $-3.3\times 10^{-3}$, whereas the $e^2$ term gives  $-3.37\times 10^{-3}$  with $K_9^r(M_\rho)+K_{10}^r(M_\rho)=1/(8\pi^2)$ and $-9.4\times 10^{-4}$ with $K_9^r(M_\rho)+K_{10}^r(M_\rho)=2.7\times 10^{-3}$, the central value  given in \cite{Bijnens:1996kk}. Therefore, the charge term above is of the same order as the pure QCD isospin correction and must be included when estimating the relative size of condensates through this sum rule. In fact, using the values quoted in \cite{Narison:2002hk}   $m_u/m_d=0.55$, $m_s/m_d=18.9$ and $\conds/\condu=0.66$, we get from (\ref{sumrule0}) with physical pion and kaon masses:

\begin{equation*}
-0.015<\frac{\condd}{\condu}-1<-0.009,
\end{equation*}
where the lower (upper) bound corresponds to the natural value $K_9^r+K_{10}^r=+(-)1/(8\pi^2)$, while the value without considering the charge correction is $-0.012$ and the value quoted in \cite{Narison:2002hk} collecting various estimates in the literature is $-0.009$. The inclusion of the charge corrections may then help to reconcile this sum rule with the different condensate estimates available. In fact, through this sum rule we see that ChPT is also compatible with the asymmetries $\condd/\condu$ and $\conds/\condu$ being both smaller than one (see our comments in section \ref{sec:resultsu3}). Note that the ferromagnetic-like arguments used in sections \ref{sec:condsu2} and \ref{sec:boundssu3} cannot be applied to $\conddif$, which does not behave as an order parameter under chiral transformations, since it is not invariant under $SU_V(2)$. Finally, we recall that estimates based on the sum rule (\ref{sumrule0}) are  more precise than the ones we have made directly from the condensates in section \ref{sec:resultsu3}, since this sum rule is free of the $H_2^r$ ambiguity.

\subsection{Matching of low-energy constants}
\label{sec:match}

Our aim in this section is to explore the consequences of including the two sources of isospin breaking for the matching of the LEC involved in the condensates. For that purpose, we perform a $1/m_s$ expansion in the $SU(3)$ sum and difference condensates given in (\ref{condsu3sum})-(\ref{condsu3dif}). Matching the $\Od(1)$ and $\Od(\log m_s)$ terms  with the corresponding $SU(2)$ expressions in (\ref{condsu2sum})-(\ref{condsu2dif}) yields the following  relations between the LEC,  for the sum and difference of condensates respectively:

\begin{widetext}

\begin{eqnarray}
  2M_{\pi^0}^2\left[l_3^r(\mu)+h_1^r(\mu)\right]+e^2 F^2 {\cal K}_2^r(\mu)&=&2M_{\pi^0}^2\left[16L_6^r(\mu)+4L_8^r(\mu)+2H_2^r(\mu)-\frac{\nu_\eta}{18}-\frac{\nu_{K^0}}{2}\right]
 \nonumber\\  &&+e^2F^2\left[{\cal K}_{3+}^r(\mu)-\frac{2C}{F^4}\nu_{K^0}\right],
\label{matchcondsum}
\end{eqnarray}
\end{widetext}

\begin{widetext}

\begin{eqnarray}
B_0(m_d-m_u)h_3-\frac{2e^2F^2}{3}k_7&=&B_0(m_d-m_u)\left[4L_8^r(\mu)+2H_2^r(\mu)-\frac{\nu_{\eta}}{3}-\frac{\nu_{K^0}}{2}+\frac{1}{96\pi^2}\right]
\nonumber\\&-&\frac{2e^2F^2}{3}\left[K_9^r(\mu)+K_{10}^r(\mu)-\frac{3C}{2F^4}\nu_{K^0}\right].
  \label{matchconddif}
\end{eqnarray}
\end{widetext}

In the above expressions, we have displayed the $SU(2)$ contribution on the left hand side and the $SU(3)$ ones on the right hand side, with ${\cal K}_2^r(\mu)$ and ${\cal K}_{3+}^r(\mu)$ given in (\ref{K2def}) and (\ref{Kscondsu3}). Note that  the $1/m_s$ expansion has been implemented also in the tree level relations (\ref{treemassessu3}), so that  $M_{\pi^0}^2=(m_u+m_d)B_0+\Od(1/m_s)$, $M_{K^0}^2= B_0 m_s+\Od(1)$ and $M_{\eta}^2= 4B_0 m_s/3+\Od(1)$. It is important to point out that the pion mass charge
difference is not negligible in the $1/m_s$
expansion, and for that reason we keep $M_{\pi^0}$ in (\ref{matchcondsum}).  For kaons, it is justified to consider the charge contribution negligible against the dominant $m_s$ term, so that at this order  $M_{K^\pm}$ and $M_{K^0}$ are not distinguishable.

In the sum  matching relation (\ref{matchcondsum}), the isospin corrections are not very significant. The mass difference $m_u-m_d$ does not appear in the neutral and kaon masses to leading order in $1/m_s$ and the charge correction, although being of the same chiral order as the $M_{\pi^0}^2$ term, numerically  $e^2 F^2/M_{\pi^0}^2\simeq C e^2/(F^2 M_{\pi^0}^2)\simeq 0.05$. However, in the difference matching (\ref{matchconddif}), the $m_u-m_d$ corrections contribute on the same footing as the EM ones and are numerically comparable.

The above matching relations can be used directly for the approximated LEC (estimated theoretically or fitted to data) and for physical masses, since the difference with the tree level masses and LEC is hidden in higher orders. On the other hand, for the tree-level LEC, i.e.  the ChPT $O(p^4)$ lagrangian parameters, since they are formally independent of the light quark masses, we can just take the chiral limit $m_u=m_d=0$ in the above expressions (\ref{matchcondsum})-(\ref{matchconddif}) and read off the corresponding matching of the $e^2$ contributions. Using the latter again in (\ref{matchcondsum})-(\ref{matchconddif}) gives then independent relations between the tree-level LEC involved at $e^2=0$ and the EM ones. Doing so, the EM and not-EM LEC combinations decouple and the results are compatible with those obtained in \cite{Gasser:1984gg} for $e=0$ and in \cite{Haefeli:2007ey} for $e\neq 0$ (setting $m_u=m_d=0$ from the very beginning):

\begin{eqnarray}
l_3^r(\mu)+h_1^r(\mu)&=&16L_6^r(\mu)+4L_8^r(\mu)+2H_2^r(\mu)-\frac{\nu_{\eta}}{18}-\frac{\nu_{K^0}}{2}\nonumber\\
h_3&=&4L_8^r(\mu)+2H_2^r(\mu)-\frac{\nu_{\eta}}{3}-\frac{\nu_{K^0}}{2}+\frac{1}{96\pi^2}\nonumber\\
5(k_5^r(\mu)+k_6^r(\mu))&=&6(K_7+K_8^r(\mu))+4(K_9^r(\mu)+K_{10}^r(\mu))-\frac{3C}{F^4}\nu_{K^0}\nonumber\\
k_7&=&K_9^r(\mu)+K_{10}^r(\mu)-\frac{3C}{2F^4}\nu_{K^0}
\label{matchingprev}
\end{eqnarray}
where the $\nu$ functions are evaluated exactly in the chiral limit, i.e, for $M_{K^0}^2= B_0 m_s$ and $M_{\eta}^2= 4B_0 m_s/3$, the first and third equation coming from (\ref{matchcondsum}) and the second and fourth from (\ref{matchconddif}).

Our first conclusion is then that to this order of approximation, the formal matching of the condensates is consistent with the matching relations previously obtained. In other words,  mass and charge terms can be separately matched. This would be no longer true at higher orders where for instance  $e^2(m_u-m_d)$ contributions may appear.

Although the relations (\ref{matchcondsum}) and (\ref{matchconddif}) reduce to (\ref{matchingprev})  in the chiral limit for the tree-level LEC,  it is better justified to use the original expressions (\ref{matchcondsum})-(\ref{matchconddif})  when dealing with  physical meson masses  and when the LEC are obtained either from phenomenological or theoretical analysis. The LEC obtained in that way are approximations to the lagrangian values and consequently they depend on mass scales characteristic of the approximation method used. For instance, the LEC obtained by  phenomenological fits are sensitive to variations both in $\hat m$ and in $m_u-m_d$ \cite{Amoros:2001cp}, in resonance saturation approaches they depend on  vector meson masses \cite{Moussallam:1997xx,Ananthanarayan:2004qk} which themselves depend on quark masses and in the NJL model some LEC like $K_{10}^r$ depend on the scale where the quark masses are renormalized \cite{Bijnens:1996kk}. We do not expect   large differences between using the general matching relation  (\ref{matchcondsum}) or  the first and third equations in  (\ref{matchingprev}), since the latter can be understood also as the $e=0$ limit of the former and  we have seen that this is numerically a good approximation. However, that is not so clear for (\ref{matchcondsum}) where the two isospin-breaking contributions are of the same order, both in the chiral expansion and numerically.

Finally, we can use the previous matching relations to estimate numerically the $SU(2)$ condensates in (\ref{condsu2sum})-(\ref{condsu2dif}) without having to appeal to the values of the $SU(2)$ LEC. Doing so we obtain $-\condsum^{SU(2)}/B_0F^2\simeq(2.16,2.18)$ and $\conddif^{SU(2)}/B_0F^2\simeq(0.014,0.02)$ where we indicate in brackets the natural range of the EM LEC, to be compared to  $-\condsum^{SU(3)}/B_0F^2\simeq(2.58,2.6)$ and $\conddif^{SU(3)}/B_0F^2\simeq(0.013,0.018)$ from Table \ref{tab:condval}. The larger difference in $\condsum$ comes from the $\Od(m_s)$ and $\Od(m_s\log m_s)$ terms in the $1/m_s$ expansion, which were separated when doing the matching and which are absent in the condensate difference. In fact, the numerical contribution of those terms to $-\condsum^{SU(3)}/B_0F^2$ is about 0.41, which explains perfectly the numerical differences and confirms the idea that in standard ChPT the light condensates calculated either in the $SU(2)$ or in the $SU(3)$ cases give almost the same answer near the chiral limit. This may be not the case in other  scenarios of chiral symmetry breaking \cite{sternmouss}.

\subsection{EM corrections and $SU(3)$ LEC bounds}
\label{sec:boundssu3}

We have seen in the $SU(2)$ case that the EM ratio
given in (\ref{emratiosu2}) is a relevant physical quantity
allowing to establish a constraint for the EM LEC based on
explicit chiral symmetry breaking. The same argument applied to
the $SU(3)$ case leads also to a constraint on the EM LEC obtained
from the full condensate $\condtwo=\langle \bar u u + \bar d d +
\bar s s\rangle$, which behaves as an order parameter, being an
isosinglet under $SU_V(3)$. In addition, we can still consider the
light condensate $\condtwo_l$ as the order parameter of chiral
transformations of the $SU(2)$ subgroup, which in principle will
lead to a different constraint. In fact, the latter is nothing but
the constraint obtained in the $SU(2)$ case (\ref{su2bound}), once
the equivalence between the LEC obtained in  section
\ref{sec:match} is used. As for the full condensate, it should be
kept in mind that the large violations of chiral symmetry due to
the strange quark mass may spoil our simple description of small
explicit breaking. As commented above, this reflects in the large
NLO contributions to the strange condensate, which in the standard
ChPT framework depends strongly on $m_s$, unlike the light
condensate. Therefore, the bounds of the LEC obtained for the full
condensate are less trustable, since neglecting higher orders, say
of $\Od(e^2 m_s)$ is not so well justified for $\conds$.
Proceeding then as in section \ref{sec:condsu2}, where the same
prescription of charge splitting when comparing with QCD
approaches is understood, we calculate the ratios:

\begin{widetext}
\begin{eqnarray}
\left[\frac{\condtwo_l^{e\neq 0}}{\condtwo_l^{e=0}}\right]^{SU(3)}
&=&1+ \frac{4e^2}{9}\left[6\left(K_7+K_8^r(\mu)\right)+5\left(K_9^r(\mu)+K_{10}^r(\mu)\right)\right]-\frac{2C e^2}{F^4}\left[2\nu_{\pi^\pm}+\nu_{K^\pm}\right]+\Od(\delta_\pi^2,\delta_K^2)+\Od(p^4)
\nonumber\\ \label{emratiosu3light} \\
\left[\frac{\condtwo^{e\neq 0}}{\condtwo^{e=0}}\right]^{SU(3)}&=&1+\frac{8}{9}e^2\left[2(K_9^r(\mu)+K_{10}^r(\mu))+3(K_7+K_8^r(\mu))\right]
-\frac{8C e^2}{3F^4}\left[\nu_{\pi^\pm}+\nu_{K^\pm}\right]+\Od(\delta_\pi^2,\delta_K^2)+\Od(p^4),\nonumber\\\label{emratiosu3full}
\end{eqnarray}
\end{widetext}
where, as in the $SU(2)$ case,  we have expanded in $\delta_\pi$
and also in
$\delta_K=(M_{K^{\pm}}^2-M_{K^{\pm},e=0}^2)/M_{K^{\pm}}^2\simeq
0.008$, which allows to express the results in terms of the full
$\pi^\pm$ and $K^\pm$ masses. Otherwise
 we should take into account that now $M_{\pi^{\pm}}(e=0)\neq M_{\pi^0}$, unlike the $SU(2)$ case, and $M_{K^{\pm}}(e=0)\neq M_{K^0}$, by terms of order $m_d-m_u$. This is important when using this result for numerical
estimates, since, as discussed before, the separation of the $e=0$
correction to the masses is formally not unique.
  As in $SU(2)$, the ratios (\ref{emratiosu3light})-(\ref{emratiosu3full}) are finite and independent of the scale $\mu$,
   of $B_0$ and of the not-EM LEC, so  they are free of contact ambiguities.

  As in section \ref{sec:condsu2},  we want to explore the consequences  of the ferromagnetic nature of the physical QCD vacuum under explicit chiral symmetry breaking for the EM LEC. Here also the charge-mass  crossed terms in the fourth order lagrangian (\ref{lag4su3}) give explicit breaking contributions to the quark condensate coming now from the isoscalar,  isovector and strangeness part of the charge matrix. For physical quark charges, demanding that the ratios (\ref{emratiosu3light})-(\ref{emratiosu3full}) are greater than one we obtain the following EM bounds, to leading order in the chiral expansion and in $\delta_\pi,\delta_K$:

\begin{widetext}
 \begin{eqnarray}
  \condtwo_l&\rightarrow& 6\left(K_7+K_8^r(\mu)\right)+5\left(K_9^r(\mu)+K_{10}^r(\mu)\right)\geq
  \frac{9C}{2 F^4}\left(2\nu_{\pi^\pm}+\nu_{K^\pm}\right)
   \label{su3boundlight}\\
   \condtwo&\rightarrow& 2\left(K_7+K_8^r(\mu)\right)+3\left(K_9^r(\mu)+K_{10}^r(\mu)\right)\geq \frac{3C}{F^4}\left(\nu_{\pi^\pm}+\nu_{K^\pm}\right)
   \label{su3boundfull}
 \end{eqnarray}
 \end{widetext}

\begin{figure}[h]
\includegraphics[scale=.55]{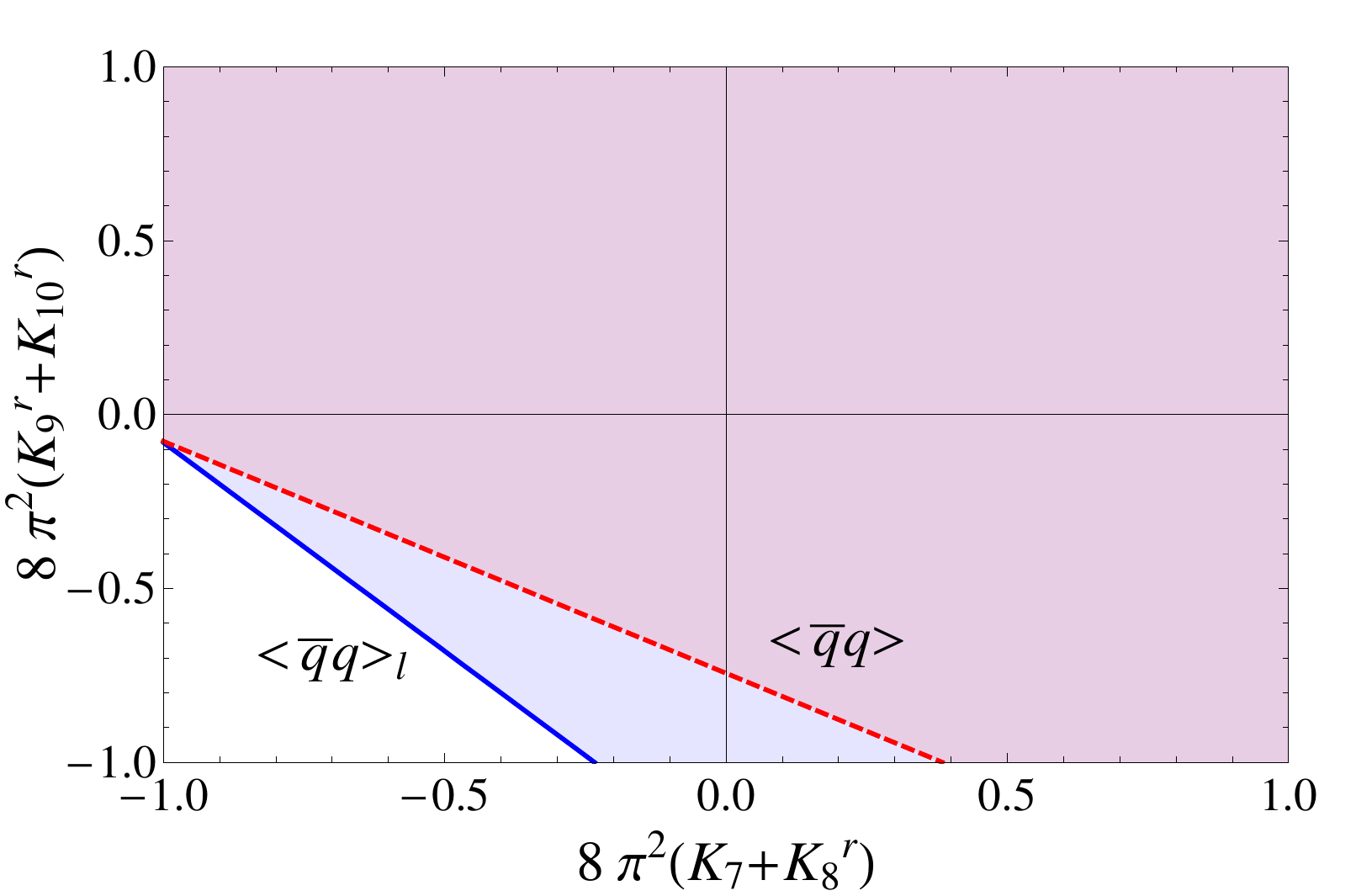}
 \caption{\rm \label{fig:bounds} Regions in the LEC space constrained by the bounds on the light condensate (\ref{su3boundlight}), above the full blue line, and the full one (\ref{su3boundfull}), above the dashed red line. The LEC are renormalized at $\mu=M_\rho$ and are plotted within the natural range.}
\end{figure}

We remark that these constraints are independent of the low-energy scale $\mu$. It is also clear that the light bound (\ref{su3boundlight}) is nothing but the one obtained in the $SU(2)$ case (\ref{su2bound}) once the equivalence between the LEC given in the third equation of (\ref{matchingprev}) is used.
 In Figure \ref{fig:bounds} we have plotted these two constraints in the $(K_7+K_8^r)-(K_9^r+K_{10}^r)$ plane at $\mu=M_\rho$ and within the natural region. We have used the same numerical values for the tree level LO masses, $C$ and $F$ as in previous sections. We observe that the bound on the full condensate is more restrictive than the light one in that range. However, as we have commented above, it is also less trustable, due to the large distortion of the chiral invariant vacuum due to the strange mass. Both bounds give also a more restrictive condition than just the natural size.

Let us now check these bounds against some estimates of the
$K_i^r$ in the literature. We start with  the sum rule approach for $K^r_{7...10}$ in
\cite{Moussallam:1997xx}. In that work, $K_7=K_8^r(M_\rho)=0$, but
what is more relevant for us  is that the combination
$K_{9}^r+K_{10}^r$ at any scale is gauge-independent despite being
both $K_9^r$, $K_{10}^r$ dependent on the gauge parameter $\xi$, as one
can readily check from the explicit expressions given in
\cite{Moussallam:1997xx} (c.f. their eqns. (94) and (95)). This is
an interesting consistency check of our present bounds
(\ref{su3boundlight}) and (\ref{su3boundfull}),  which are gauge independent in addition to their $\mu$ low-energy scale independency commented previously, supporting their validity and predictive character. Numerically,
the constant $K_9^r$ could not be estimated in
\cite{Moussallam:1997xx} due to the slow convergence of the
integrals involved, but they  provide a numerical estimate for
$K_{10}^r(M_\rho)=5.2 \times 10^{-3}$ at $\mu_0=0.7$ GeV and $\xi=1$, for which we get $K_9^r(M_\rho)\geq -0.021$ from (\ref{su3boundlight}) and
$K_9^r(M_\rho)\geq -0.015$ from (\ref{su3boundfull}). See our comments about the $\mu_0$ scale dependence in section \ref{sec:condsu2}.

 In \cite{Baur:1996ya} resonance saturation gives $K^r_{7...10}(M_\rho)=0$,
which is compatible with our present bound. This is apparently
incompatible with a previous value for $K_8^r(M_\rho)=(-4\pm
1.7)\times 10^{-3}$ obtained  in \cite{Urech:1994hd}.
The possible reasons to explain this difference were discussed in
\cite{Baur:1996ya}. That value for $K_8^r$ is compatible
with our bounds as long as $6K_7+5(K_9^r+K_{10}^r)\geq -0.05$ from (\ref{su3boundlight}) and $2K_7+3(K_9^r+K_{10}^r)\geq -0.02$ from (\ref{su3boundfull}).

In \cite{Bijnens:1996kk},
based on large $N_c$ and the NJL-model, the LEC estimates give $K_7=0$,
$K_8(M_\rho)=(-0.8\pm 2.0)\times 10^{-3}$ ($K_7$ and $K_8$ are
$\Od(1/N_c)$ suppressed) and $K_9^r(M_\rho)+K_{10}^r(M_\rho)=(2.7\pm
1.0)\times 10^{-3}$, all of them at $\mu_0=0.7$  GeV. These values are also compatible with both
bounds (\ref{su3boundlight}) and (\ref{su3boundfull}). We note that in \cite{Pinzke:2004be}, where the short-distance contributions are evaluated
 as in \cite{Bijnens:1996kk}, the  explicit expressions given for the LEC  show again that $K_7,K_8^r$ and $K_9^r+K_{10}^r$ are gauge independent. Furthermore, $K_9^r$ and $K_{10}^r$,  dominant for large $N_c$ in that approach,  show a rather large stability range in the $\mu_0$ scale around $\mu_0=0.7$ GeV \cite{Bijnens:1996kk,Pinzke:2004be}. Since $K_9^r+K_{10}^r$ is the only combination  surviving for large $N_c$ in our bounds, the comparison with those works is robust concerning the gauge and QCD scale  dependence.

\section{Conclusions}

In this work we have carried out an analysis of  strong and
electromagnetic  isospin-breaking corrections to the quark
condensates in standard one-loop ChPT, providing their explicit expressions and studying  some of their main phenomenological consequences for two and three light flavours.

Our results have allowed us to analyze the sensitivity of recent isospin-breaking numerical analysis of the condensates to considering all the EM LEC involved.  The effect of those LEC is  smaller for
 individual condensates than for the vacuum asymmetry, where they show up already in the
  leading order. These corrections lie within the error range quoted in   lattice analysis. Our analysis   can also be used to estimate  corrections to the quark condensate coming from lattice artificially large meson masses.

We have shown that if EM explicit chiral symmetry breaking  induces a ferromagnetic-like response of the physical QCD vacuum, as in the case of
 quark masses, one obtains useful constraints as lower bounds for certain combinations of the EM LEC, both in the two and three flavour sectors. We have explored the consequences of this behaviour for the ratios of $e\neq 0$ to $e=0$ light and total quark condensates, which are free of contact-term ambiguities, and for a given convention of charge separation. The
  large ChPT corrections to the strange condensate make the constraints on the full condensate less reliable.
 In this context, we have discussed the different sources for  EM explicit chiral symmetry breaking  and isospin breaking terms, by considering formally arbitrary quark charges. Thus, there are chiral symmetry breaking terms proportional to the sum of charges squared, coming from crossed charge-mass contributions in the effective action, which show up in the vacuum expectation value. In accordance with the external source method, we keep the quark masses different from zero  to account correctly for all the explicit symmetry breaking sources. The chiral limit can be taken at the end of the calculation.  The bounds obtained are explicitly independent of the low-energy scale $\mu$,  providing then a complete and model-independent prediction at low energies. However, when this low-energy representation is compared with theoretical estimates  based on QCD, one has to take into account that due to the convention used in the charge separation,  the estimated LEC depend on  the QCD renormalization scale $\mu_0$, as well as being gauge dependent. Our bounds are numerically compatible with those estimates, based on sum rules, resonance saturation and QCD-like models, within the stability range of $\mu_0$ where those approaches are reliable. Furthermore, the LEC combinations appearing in our bounds are gauge independent.  We believe that our results can be useful in view of
the few estimates of the EM LEC in the literature.

We have found that the EM correction to the
sum rule relating condensate ratios is of the same order as the previously calculated $e=0$ one,
and therefore must be taken into
account when using this sum rule to estimate the relative size of
quark condensates. We have actually showed that using the complete sum rule, which is also free of contact terms,  yields  a ChPT model-independent prediction for the vacuum asymmetry compatible with the
results quoted in the literature.

Finally, we have performed a matching between the $SU(2)$ and $SU(3)$ condensates, including all isospin-breaking terms. Matching the sum and difference of light condensates gives rise to matching relations between the LEC involved, where EM and not-EM LEC enter on the same footing in the chiral expansion. These matching relations may be useful when working with physical masses and LEC estimated by different approximation methods. In the case of the sum,  the charge contribution is numerically small with respect to the pion mass one, but in the difference   the two sources of isospin-breaking are comparable.  Taking the chiral limit,   EM and not-EM constants decouple and the matching conditions are compatible with previous works for the LEC in the lagrangian, which are defined in this limit.

\appendix

\section{Fourth order lagrangians and renormalization of the LEC}
\label{app:lag}

We collect here some results available in the literature and needed in the main text. To calculate the quark condensates to NLO one needs the ${\cal L}_{p^4+p^2e^2+e^4}$ lagrangians to absorb the divergences coming from loops with vertices from  ${\cal L}_{p^2+e^2}$. We denote by a superscript $\bar q q$ the relevant terms in the lagrangian, which are those containing the quark mass matrix. For $SU(2)$ they are \cite{Meissner:1997fa,Knecht:1997jw}:

\begin{widetext}

\begin{equation}\label{lag4su2}
\begin{split}
\mathcal{L}_{p^4}^{\bar qq}&=\frac{l_3}{16}\tr[ \chi(U+U^\dagger)]^2+\frac{1}{4}(h_1+h_3)\tr[\chi^2]+\frac{1}{2}(h_1-h_3) \det(\chi),
\\
\mathcal{L}_{p^2e^2}^{\bar qq}&= F^2 \bigg(k_5\tr[\chi(U+U^\dagger)] \tr[ Q^2]
    +k_6\tr[\chi(U+U^\dagger)]\tr[QUQU^\dagger]
    +k_7 \tr[(\chi U^\dagger+U\chi)Q +(\chi U+U^\dagger\chi)Q]\tr[Q]\bigg),
    \end{split}
\end{equation}
\end{widetext}

where $\chi=2B_0{\cal M}$, whereas for SU(3) \cite{Urech:1994hd}:

\begin{widetext}
\begin{gather}
\begin{split}
\mathcal{L}_{p^4}^{\bar qq}=&L_6\tr[\chi(U+ U^\dagger)]^2+L_8\tr[\chi U\chi U+\chi U^\dagger \chi U^\dagger]+  H_2\tr[\chi^2]
\\
\mathcal{L}_{p^2e^2}^{\bar qq}=& F^2 \bigg(K_7\tr[\chi(U+U^\dagger)]\tr[Q^2]+K_8\tr[\chi(U+U^\dagger)] \tr[ QUQU^\dagger]+K_9\tr[\left(\chi U+U^\dagger\chi+\chi U^\dagger+U\chi\right)Q^2]+\\& K_{10}\tr\left[\left(\chi U+U^\dagger\chi\right)QU^\dagger QU\right.+\left.\left(\chi U^\dagger+ U\chi\right)QUQU^\dagger\right]\bigg),
\end{split}
\label{lag4su3}
\end{gather}
\end{widetext}
and $\mathcal{L}_{e^4}^{\bar qq}=0$ for both cases.

In order to renormalize the meson loops it is necessary to separate the low energy constants appearing in the NNLO lagrangian in finite and divergent parts.
The renormalization of the LEC involved in the calculation of the SU(2) condensates is given by \cite{Gasser:1983yg,Meissner:1997fa,Knecht:1997jw}:
\begin{gather*}
\begin{split}
l_i=& l^r_i(\mu)+\gamma_i \lambda,
\\
h_i=& h^r_i(\mu)+\delta_i \lambda,
\\
k_i=& k_i^r(\mu)+\sigma_i \lambda,
\end{split}
\end{gather*}
with $\gamma_3=-\frac{1}{2}$, $\delta_1=2$, $\delta_3=0$, and
$\sigma_5=-\frac{1}{4}-\frac{1}{5}Z$, $\sigma_6=\frac{1}{4}+2Z$
and $\sigma_7=0$, for physical quark charges $e_u=2e/3$,
$e_d=-e/3$, being $Z:=\frac{C}{F^4}$. The part that diverges in
$d=4$ dimensions is isolated from the loop integrals and is
expressed as
$$\lambda=\frac{\mu^{d-4}}{16\pi^2}\left(\frac{1}{d-4}-\frac{1}{2}[\log4\pi+\Gamma'(1)+1]\right),$$
where $-\Gamma'(1)$ is the Euler constant. As for the SU(3) ones, we have \cite{Gasser:1984gg,Urech:1994hd}
\begin{gather*}
\begin{split}
L_i=& L^r_i(\mu)+\Gamma_i \lambda,
\\
H_i=& H^r_i(\mu)+\Delta_i \lambda,
\\
K_i=& K_i^r(\mu)+\Sigma_i \lambda,
\end{split}
\end{gather*}
with $\Gamma_6=\frac{11}{144}$, $\Gamma_8=\frac{5}{48}$, $\Delta_2=\frac{5}{24}$, and
 $\Sigma_7=0$, $\Sigma_8=Z$, $\Sigma_9=-\frac{1}{4}$, $\Sigma_{10}=\frac{1}{4}+\frac{3}{2}Z$.

\section*{Acknowledgments}
We are grateful to J.R.Pel\'aez and E.Ruiz Arriola for useful comments. R.T.A would like to thank Buenaventura Andr\'es L\'opez for invaluable advice. Work partially supported by the Spanish
research contracts FPA2008-00592,  FIS2008-01323, UCM-Santander 910309  GR58/08, GR35/10-A  and the FPI programme (BES-2009-013672).

\end{document}